\documentclass[12pt]{iopart}

\usepackage{siunitx}
\usepackage{graphicx}
\usepackage{cleveref}
\usepackage{tabularx}
\usepackage{bigstrut}
\usepackage{floatrow}
\usepackage{longtable}
\usepackage{xcolor}
\usepackage{colortbl}
\usepackage{multirow}
\usepackage{booktabs}
\usepackage{multicol}
\usepackage{color}
\usepackage{url}

\newfloatcommand{capbtabbox}{table}[][\FBwidth]
\usepackage[caption = false]{subfig}
\graphicspath{Figures}
\begin{document}
\title[]{Satellite 
constellations for trusted node QKD networks}

\author{T Vergoossen$^1$\footnote{Author, Satellite-QKD model.}, S Loarte$^{1,2}$\footnote{Constellation model and analysis in this work.},  R Bedington$^1$, H Kuiper$^2$ and A Ling$^{1,3}$}\address{$^1$ Centre for Quantum Technologies, National University of Singapore, 3 Science Drive 2, Singapore 120435}\address{$^2$ Faculty of Aerospace Engineering, Delft University of Technology, Kluyverweg 1, 2629 HS Delft, The Netherlands}\address{$^3$ Department of Physics, National University of Singapore, 2 Science Drive 3, Singapore 117551}\ead{cqttv@nus.edu.sg}

\vspace{10pt}
\begin{indented}
\item[]March 2019
\end{indented}

\begin{abstract}
Quantum key distribution from satellites becomes particularly valuable when it can be used on a large network and on-demand to provide a symmetric encryption key to any two nodes. A constellation model is described which enables QKD-derived encryption keys to be established between any two ground stations with low latency. This is achieved through the use of low earth orbit, trusted-node QKD satellites which create a buffer of keys with the ground stations they pass over, and geostationary relay satellites to transfer secure combinations of the keys to the ground stations. Regional and global network models are considered and the use of inter-satellite QKD links for balancing keys is assessed.
\end{abstract}

%
\vspace{2pc}

\section{Introduction} 

Quantum key distribution (QKD) is a branch of quantum cryptography which describes methods for establishing highly secure symmetric keying material between separated users \cite{Gisin2002}. QKD processes use very weak optical signals so have fundamental distance limitations due to increased losses with increasing distances. To extend the separations possible between users repeater nodes can be used, either `trusted nodes' or quantum repeaters. The most secure solution is to use quantum repeaters; these do not make a measurement on the QKD signals so the communicating parties are able to verify between themselves that the key they have established is secure. Quantum repeaters however have yet to be demonstrated in practical systems and are not ready to be considered for near term QKD networks. Present QKD networks accordingly use trusted nodes. In satellite QKD users establish keys with the satellite node, which then combines their keys as an XOR (exclusive OR operation) and broadcasts this publicly \cite{Bedington2017,Liao2018}. The XOR key is meaningless to anyone except the two users who can use it to determine each others keys, which they can then use as secret key material for encryption purposes. In optical fibre networks, many trusted nodes can be connected together to create long connections such as the Beijing-Shanghai QKD link, which features 32 trusted nodes \cite{Courtland2016}. Since these nodes have full access to the keys passing through them they must be trusted to be secure, and the security can only come from conventional security methods---e.g. restricted access and trusted human guards. As a result, links with large numbers of nodes, and nodes based in foreign countries, are intrinsically harder to trust. Low earth orbit (LEO) satellite-based trusted nodes help address both of these issues by reducing the numbers of nodes required. Firstly, LEO satellites orbit the Earth continuously and pass over most places several times per day---rather than using a chain of nodes they can simply act as a store-and-forward device and wait until the desired recipient passes underneath them \footnote{QKD is a point to point service and it is assumed that the keying material is buffered for future use so the low latency of the store-and-forward technique is not an issue.}. Secondly, they are likely to be harder to access and infiltrate than any ground-based facility.         

Satellite QKD has been extensively discussed in terms of theoretical modelling descriptions \cite{Bourgoin2012}, technology developments \cite{Steinlechner2012,Naughton2019}, hardware demonstrations \cite{Takenaka2017,Vallone2015,Grieve2018,Liao2018} and in review articles \cite{Bedington2017,Kahn2018}. It is clear, and often stated, that the logical successor to single QKD satellites is a QKD constellation \cite{Skander2010,Elser2012}, but how this could be implemented has yet to be investigated or defined in published literature. In this study we define a concept for a low earth orbit (LEO) trusted node QKD satellite and investigate its effectiveness in different constellation arrangements. In particular we assess the value of inter-satellite QKD links.


\section{Methods} 
We describe a model capable of analysing the performance of satellite networks featuring satellite-to-ground and inter-satellite links, and a concept for a trusted node QKD constellation.

\subsection{Satellite-QKD Links}
Satellite-QKD links are modelled in Matlab using principles set forth by Bourgoin et al. \cite{Bourgoin2012}. Key rates are calculated from optical link parameters using the equations for weak and vacuum decoy-state BB84 provided by Xiongfeng Ma et al. \cite{Ma2005}. To validate the model, the QKD demonstrations performed by the Micius satellite were modelled and compared to the published results (input parameters are in Appendix A) \cite{Liao2017}. \Cref{fig:miciusvalidation} shows that there is a lot more variability in the real-world results, but that the model results are placed convincingly within this variability.

Some simplifications of the model include assuming a constant pointing error and constant, but path length dependent, atmospheric losses rather than experiencing local weather effects. Furthermore the asymptotic key size assumption is used and intensity fluctuations of signal and decoy states are not considered. As a result error correction and privacy amplification losses are underestimated.

 \begin{figure}[H]
\centering
\subfloat{\includegraphics[width = \textwidth]{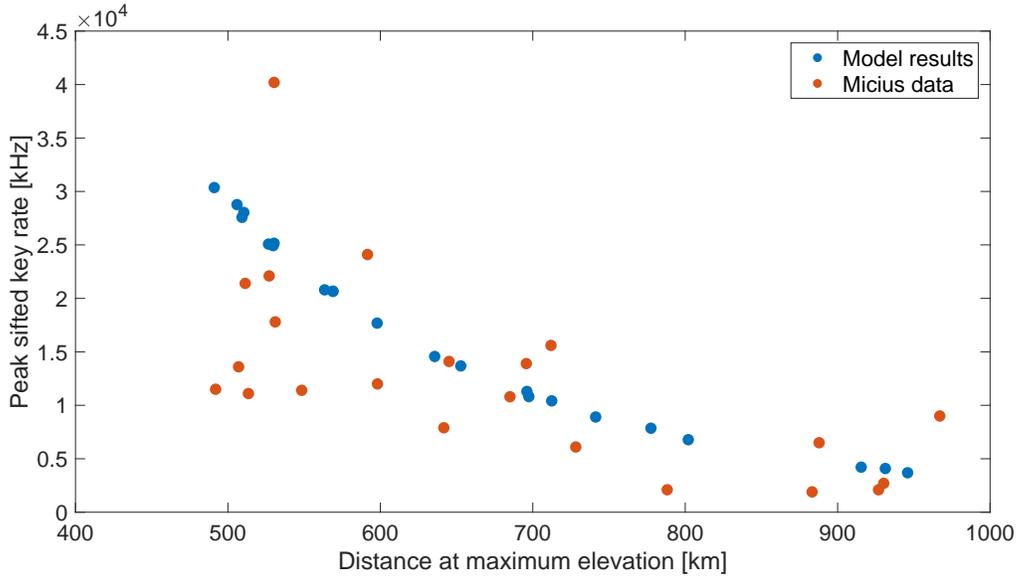}}
\centering
\caption{Comparison of our model results with Micius published data for specific passes between September 2016 and May 2017.}
\label{fig:miciusvalidation}
\end{figure}

\subsection{QKD trusted node constellation concept}
In our constellation concept, satellites perform QKD with any ground station they pass over to generate a private symmetric key between that satellite and the ground station. Each satellite maintains a continuous classical radio link with the ground network, using relay nodes in higher orbits, e.g. in geostationary orbit \cite{addvalue}. When any two users on the ground (e.g. ground station A and B), need to communicate securely the network is queried for satellites that have keys shared with A and B on-board a secure key management module. Any such keys are combined using an XOR function and can be broadcast to A and B via the relay satellites with minimal latency. Station A and B can perform a reverse XOR operation to extract the key held by the other ground station and can now communicate securely. The used key for A and B is deleted from satellites.

We also propose an extension to the above configuration in which the LEO satellites have quantum inter-satellite links (ISL) that are used to re-distribute key across the constellation. This solves the problem where some satellites in the constellation have secure key established with station A, but none or not as much with station B; and vice versa. In this situation these satellites either cannot create an XOR of key A and B, or they are constrained by the size of one of the keys. However, secure re-distribution of keys A and B between satellites would maximise the mutual key available on each satellite for XOR operations and optimise the constellation effectiveness. Inter-satellite links are feasible if the satellites are within close proximity of each other and have low relative angular motion. These operational requirements lead to two possible scenarios: ISL between satellites in different orbital planes (inter-planar) and a string-of-pearls configuration of satellites in the same plane (intra-planar ISL). This is illustrated in \Cref{fig:resultsandISL}d. In a string-of-pearls configuration satellites would have additional receivers/transmitters to perform QKD with satellites ahead of them or behind them, while the inter-planar case requires more capable link architectures to communicate over larger relative angles and distances. In addition, intra-planar links provide constant availability, provided they are in eclipse and so will outperform the inter-planar links. For these reason only the string-of-pearls configuration is discussed further in this paper.

\subsection{Constellation Modelling}
\begin{figure}[H]
  \centering
  \includegraphics[width=\textwidth]{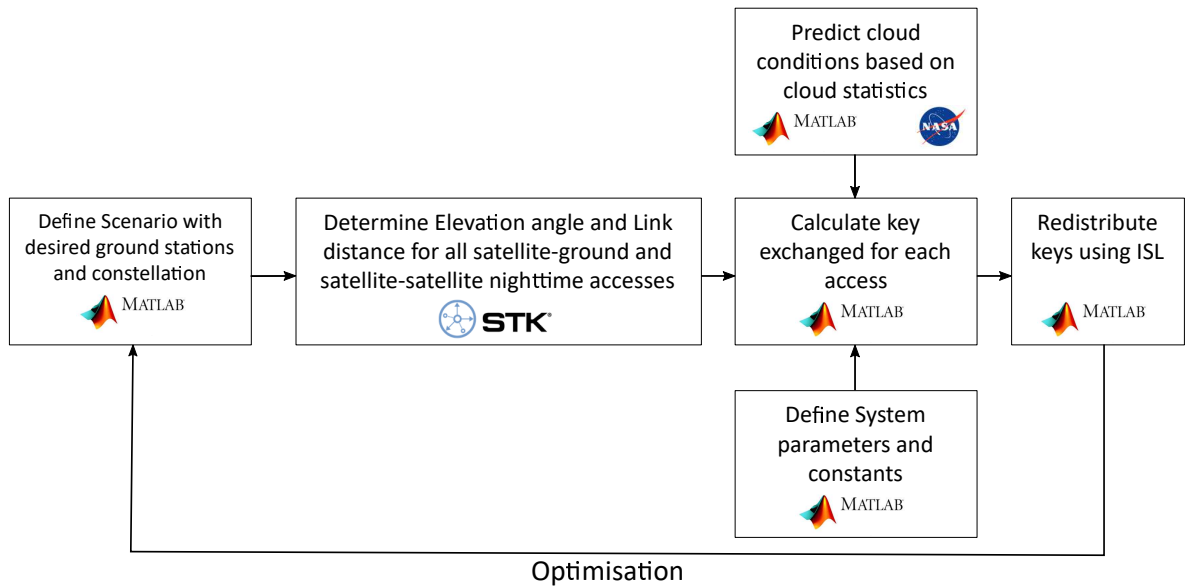}
  \caption{Architecture of the toolkit for modelling QKD constellations}
  \label{fig:modelarchitecture}
 \end{figure}

The satellite QKD Matlab model was integrated with STK's orbit modelling capabilities and extended to model constellations with satellite-to-ground and inter-satellite links. Additionally, cloud coverage predictions per pass are calculated based on historical cloud statistics \cite{cloudstatistics}. The model architecture is shown in \Cref{fig:modelarchitecture}. 

The terminals are assumed to only be able to perform QKD at night so the model is configured to perform QKD only during passes where both the satellite and the ground station are in eclipse. They are assumed to convert and store sufficient solar power in the rest of the orbit that they can perform QKD continuously when in eclipse. Satellites have one Micius-type Earth-pointing terminal, and where specified another terminal available for inter-satellite links. In the case of overlapping ground station passes, i.e. when ground stations are close together and are simultaneously in view of the satellite, the satellite will choose to perform QKD with the less cloudy one. 

Two example ground station networks are considered in this analysis to demonstrate how global and regional networks can have different considerations. The global example uses nodes in the G20 cities (Brussels is considered to serve all Western European cities), the regional example is of an Indo-ASEAN network, see appendix B. The satellite networks only consist of quasi-circular orbits and combinations of orbital planes at the same altitude.

\section{Results}
\begin{figure}[t!p]
\subfloat[]{
            \centering
            \includegraphics[width=.5\textwidth]{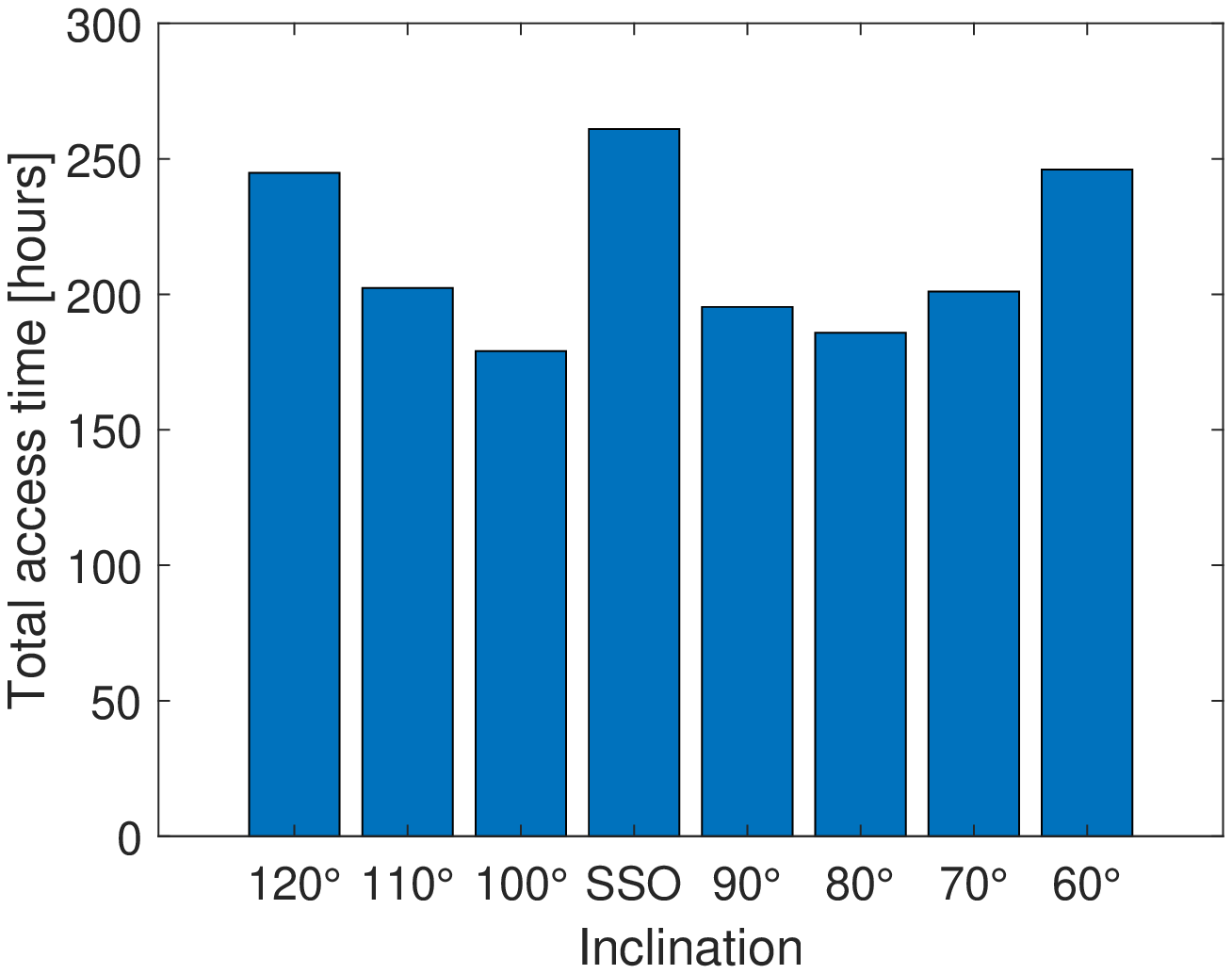}
            \label{fig:G20_single}
        }
\subfloat[]{
            \centering
            \includegraphics[width=.5\textwidth]{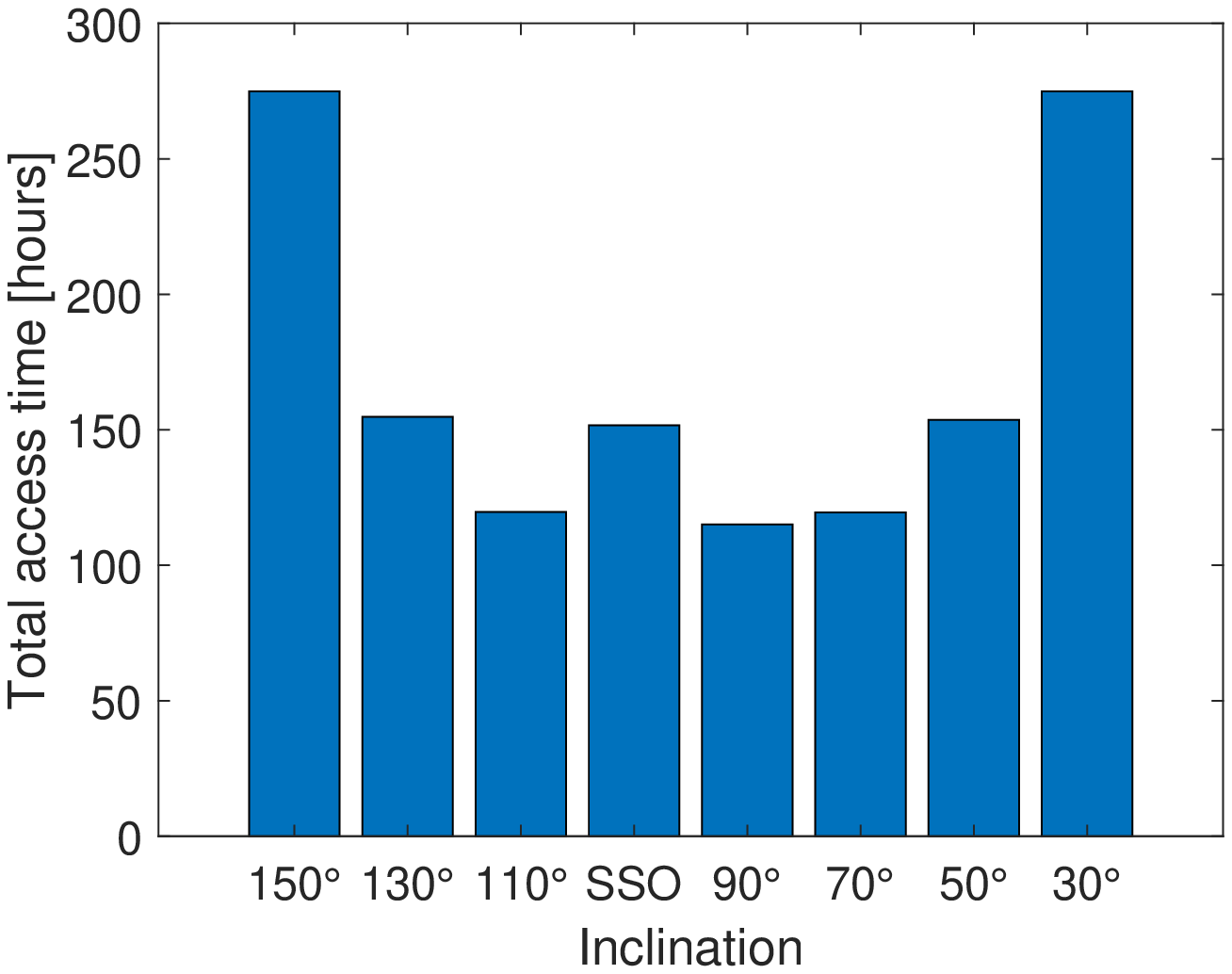}
            \label{fig:IA_single}
        }

\subfloat[]{
             \centering
             \includegraphics[width=.5\textwidth]{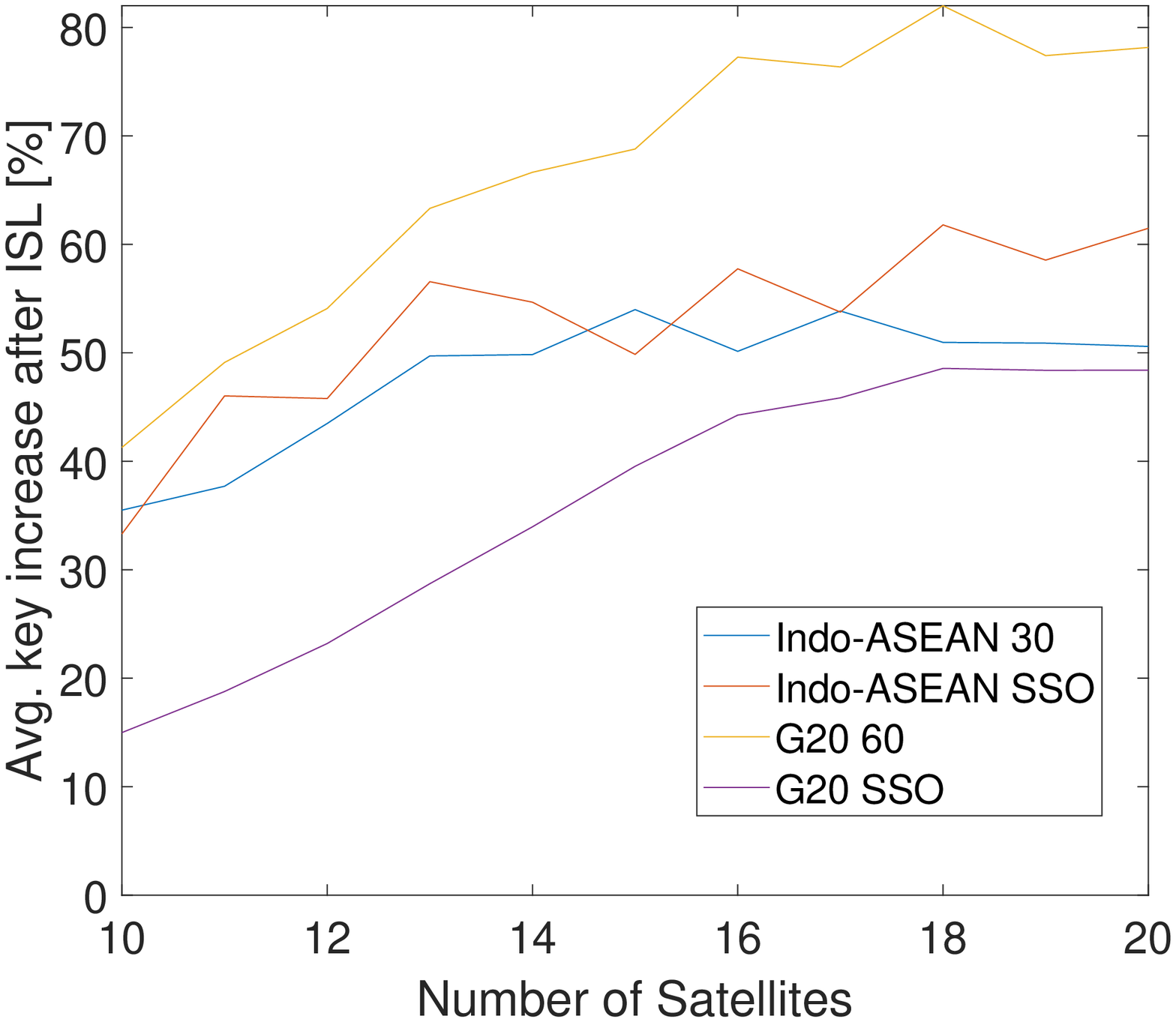}
             \label{fig:ISL_optim}
       }
\subfloat[]{
             \centering
             \includegraphics[width=.55\textwidth]{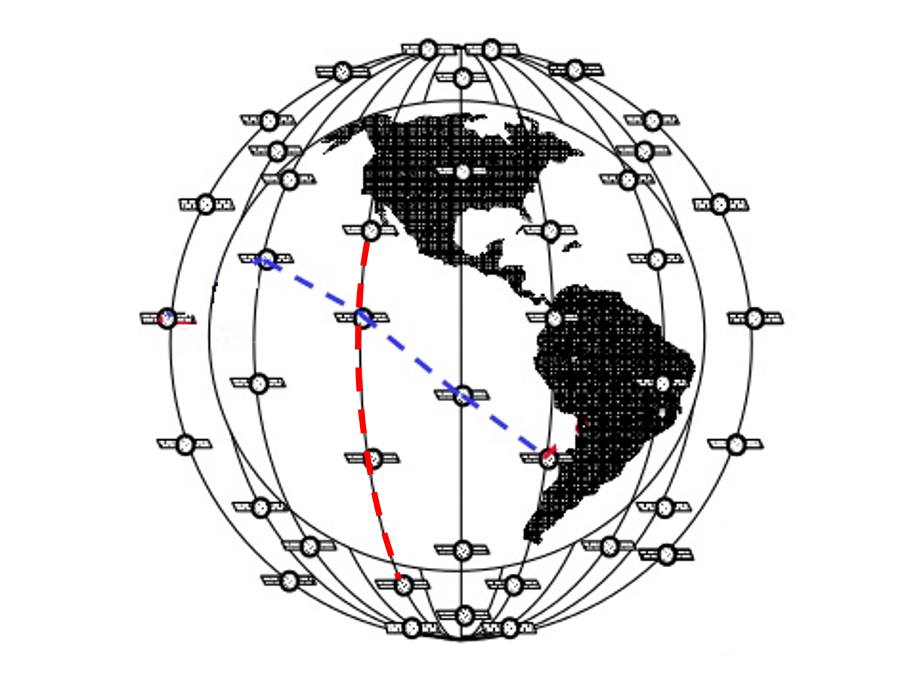}
              \label{fig:ISL_types}
       }
\caption{
(a) Total access time of 6 satellites equally distributed in a single-plane over one year for the G20 network. SSO refers to a noon-midnight Sun-Synchronous Orbit and corresponds to an inclination of 96-100 degrees depending on altitude (b) for the Indo-ASEAN network (c) Diminishing return of increasing number of satellites with ISL in constellation leads to optimum size of constellation. Essentially, there is no need for satellite to share keys in excess of the key material they need to re-distribute. (d) Possible types of ISL. Only intra-planar (red) links are considered in this paper as they are less complex and have higher link availability.}  \label{fig:resultsandISL}
\end{figure}

First, total access times for single-plane constellations at different inclination angles were evaluated in \Cref{fig:resultsandISL}a and b. Second, key rates for multi-plane constellations were investigated where all planes have the same inclination angle but are spaced evenly around the globe.

There are many different potential use cases for QKD networks, and thus many different ways to optimise and compare implementations. In this study we use as our figure of merit, the maximum message size that could be sent out to all of the other stations in the network using one-time-pad (bit for bit encrypting) and using each key only once. This `embassy model' is the scenario for example if a country wanted to send out the same secret message to all of its overseas embassies. 

\Cref{tab:G20_results} and \Cref{tab:IA_results} show the maximum distributed messages nodes can send for different constellation configurations. In practice this is limited by how much key other nodes hold. The default number of satellites used for simulations is six, and the ISL configurations use the optimal number of satellites as established in \Cref{fig:resultsandISL}c. 


\begin{table}[htbp]
  \centering
  \caption{Distributed message size after 1 year for the G20 network of ground stations for constellations of 6 and 16 satellites. For example, over one year Ankara could send 49.55 Mbits of secure messages to all cities in this list, provided the cities have enough key of their own. SSO refers to a noon-midnight Sun-Synchronous Orbit. The notation [60 deg 2p] indicates that the constellation is distributed equally over 2 orbital planes with an inclination of 60 degrees. Within one plane the argument of periapsis of the 3 satellites is 360/3 = 120 degrees apart. The longitude of the ascending node of the orbital planes themselves is 360/2=180 degrees apart. The colours are a visual aid for identifying key sizes (from red to green).}
    \begin{tabular}{|p{6em}|c|c|c|c|c|cc|cc|}
    \hline
    \textbf{G20} & \multicolumn{4}{p{11em}}{6 satellites} & \multicolumn{1}{c}{} & \multicolumn{4}{p{11.29em}|}{16 satellites} \bigstrut\\
    \hline
    \multicolumn{1}{|r|}{} & \multicolumn{4}{p{11em}|}{ Message size [Mbit]} &       & \multicolumn{4}{p{11.29em}|}{ Message size [Mbit]} \bigstrut\\
\cline{2-5}\cline{7-10}    Location & \multicolumn{1}{p{2.75em}|}{SSO } & \multicolumn{1}{p{2.75em}|}{60deg 2p/3s} & \multicolumn{1}{p{2.75em}|}{60 deg 3p/2s} & \multicolumn{1}{p{2.75em}|}{60 deg 16p/1s} &       & \multicolumn{1}{p{2.75em}|}{SSO 1p/16s} & \multicolumn{1}{p{2.75em}|}{SSO  1p/16s ISL} & \multicolumn{1}{p{2.75em}|}{60 deg } & \multicolumn{1}{p{3.04em}|}{60deg 1p/6s ISL} \bigstrut\\
\hline    Ankara & \cellcolor[rgb]{ .753,  .851,  .506}49.55 & \cellcolor[rgb]{ .922,  .902,  .514}40.86 & \cellcolor[rgb]{ .969,  .914,  .518}38.47 & \cellcolor[rgb]{ .937,  .906,  .518}39.97 &       & \cellcolor[rgb]{ .996,  .902,  .518}129.73 & \cellcolor[rgb]{ .392,  .749,  .486}182.76 & \cellcolor[rgb]{ .98,  .831,  .518}102.43 & \cellcolor[rgb]{ .714,  .839,  .502}157.91 \bigstrut[t]\\
    Beijing & \cellcolor[rgb]{ .898,  .894,  .514}42.01 & \cellcolor[rgb]{ .992,  .89,  .518}34.48 & \cellcolor[rgb]{ .996,  .902,  .518}35.37 & \cellcolor[rgb]{ .992,  .89,  .518}34.38 &       & \cellcolor[rgb]{ .984,  .843,  .518}106.62 & \cellcolor[rgb]{ .549,  .792,  .494}170.66 & \cellcolor[rgb]{ .976,  .812,  .518}94.54 & \cellcolor[rgb]{ .714,  .839,  .502}157.91 \\
    Brasilia & \cellcolor[rgb]{ .902,  .894,  .514}41.81 & \cellcolor[rgb]{ .996,  .922,  .518}37.00 & \cellcolor[rgb]{ .996,  .918,  .518}36.58 & \cellcolor[rgb]{ .969,  .914,  .518}38.55 &       & \cellcolor[rgb]{ .984,  .851,  .518}109.64 & \cellcolor[rgb]{ .392,  .749,  .486}182.76 & \cellcolor[rgb]{ .98,  .831,  .518}102.06 & \cellcolor[rgb]{ .714,  .839,  .502}157.91 \\
    Brussels (Europe) & \cellcolor[rgb]{ .969,  .761,  .518}23.82 & \cellcolor[rgb]{ .976,  .8,  .518}26.97 & \cellcolor[rgb]{ .976,  .796,  .518}26.72 & \cellcolor[rgb]{ .976,  .812,  .518}27.92 &       & \cellcolor[rgb]{ .961,  .725,  .518}61.68 & \cellcolor[rgb]{ .992,  .886,  .518}122.57 & \cellcolor[rgb]{ .969,  .753,  .518}72.53 & \cellcolor[rgb]{ .902,  .894,  .514}143.71 \\
    Bueno Aires & \cellcolor[rgb]{ .831,  .875,  .51}45.48 & \cellcolor[rgb]{ .957,  .91,  .518}39.02 & \cellcolor[rgb]{ .969,  .914,  .518}38.44 & \cellcolor[rgb]{ .945,  .906,  .518}39.58 &       & \cellcolor[rgb]{ .992,  .882,  .518}121.02 & \cellcolor[rgb]{ .388,  .745,  .482}182.76 & \cellcolor[rgb]{ .98,  .824,  .518}99.46 & \cellcolor[rgb]{ .714,  .839,  .502}157.91 \\
    Canberra & \cellcolor[rgb]{ .867,  .886,  .514}43.61 & \cellcolor[rgb]{ .953,  .91,  .518}39.31 & \cellcolor[rgb]{ .969,  .914,  .518}38.44 & \cellcolor[rgb]{ .988,  .922,  .518}37.43 &       & \cellcolor[rgb]{ .988,  .875,  .518}118.44 & \cellcolor[rgb]{ .4,  .749,  .486}182.13 & \cellcolor[rgb]{ .984,  .835,  .518}104.22 & \cellcolor[rgb]{ .714,  .839,  .502}157.91 \\
    Delhi & \cellcolor[rgb]{ .89,  .89,  .514}42.50 & \cellcolor[rgb]{ .988,  .918,  .518}37.54 & \cellcolor[rgb]{ .996,  .91,  .518}36.10 & \cellcolor[rgb]{ 1,  .922,  .518}36.79 &       & \cellcolor[rgb]{ .988,  .859,  .518}112.53 & \cellcolor[rgb]{ .451,  .765,  .486}178.17 & \cellcolor[rgb]{ .98,  .82,  .518}97.34 & \cellcolor[rgb]{ .714,  .839,  .502}157.91 \\
    Jakarta & \cellcolor[rgb]{ .973,  .776,  .518}24.89 & \cellcolor[rgb]{ .965,  .745,  .518}22.43 & \cellcolor[rgb]{ .965,  .745,  .518}22.40 & \cellcolor[rgb]{ .965,  .749,  .518}22.94 &       & \cellcolor[rgb]{ .965,  .737,  .518}66.49 & \cellcolor[rgb]{ .996,  .906,  .518}131.33 & \cellcolor[rgb]{ .961,  .718,  .518}59.02 & \cellcolor[rgb]{ .988,  .875,  .518}118.03 \\
    Mexico City & \cellcolor[rgb]{ .706,  .839,  .502}51.99 & \cellcolor[rgb]{ .82,  .871,  .51}46.07 & \cellcolor[rgb]{ .8,  .867,  .51}47.02 & \cellcolor[rgb]{ .796,  .863,  .506}47.28 &       & \cellcolor[rgb]{ .906,  .898,  .514}143.21 & \cellcolor[rgb]{ .392,  .749,  .486}182.76 & \cellcolor[rgb]{ .992,  .89,  .518}124.05 & \cellcolor[rgb]{ .714,  .839,  .502}157.91 \\
    Moscow & \cellcolor[rgb]{ .957,  .69,  .518}17.83 & \cellcolor[rgb]{ .976,  .8,  .518}26.90 & \cellcolor[rgb]{ .973,  .788,  .518}25.85 & \cellcolor[rgb]{ .976,  .804,  .518}27.24 &       & \cellcolor[rgb]{ .957,  .69,  .518}47.05 & \cellcolor[rgb]{ .976,  .812,  .518}94.10 & \cellcolor[rgb]{ .969,  .761,  .518}74.31 & \cellcolor[rgb]{ .871,  .886,  .514}146.06 \\
    Ottawa & \cellcolor[rgb]{ .988,  .867,  .518}32.51 & \cellcolor[rgb]{ .984,  .839,  .518}30.11 & \cellcolor[rgb]{ .984,  .851,  .518}31.10 & \cellcolor[rgb]{ .98,  .82,  .518}28.59 &       & \cellcolor[rgb]{ .976,  .796,  .518}87.89 & \cellcolor[rgb]{ .941,  .906,  .518}140.69 & \cellcolor[rgb]{ .969,  .773,  .518}79.39 & \cellcolor[rgb]{ .745,  .851,  .506}155.66 \\
    Pretoria & \cellcolor[rgb]{ .659,  .824,  .498}54.26 & \cellcolor[rgb]{ .773,  .859,  .506}48.42 & \cellcolor[rgb]{ .769,  .855,  .506}48.69 & \cellcolor[rgb]{ .792,  .863,  .506}47.47 &       & \cellcolor[rgb]{ .902,  .894,  .514}143.50 & \cellcolor[rgb]{ .392,  .749,  .486}182.76 & \cellcolor[rgb]{ .992,  .894,  .518}125.99 & \cellcolor[rgb]{ .714,  .839,  .502}157.91 \\
    Riyadh & \cellcolor[rgb]{ .388,  .745,  .482}68.12 & \cellcolor[rgb]{ .561,  .796,  .494}59.46 & \cellcolor[rgb]{ .561,  .796,  .494}59.39 & \cellcolor[rgb]{ .573,  .8,  .494}58.74 &       & \cellcolor[rgb]{ .392,  .749,  .486}182.76 & \cellcolor[rgb]{ .392,  .749,  .486}182.76 & \cellcolor[rgb]{ .714,  .839,  .502}157.91 & \cellcolor[rgb]{ .714,  .839,  .502}157.91 \\
    Seoul & \cellcolor[rgb]{ .98,  .918,  .518}37.91 & \cellcolor[rgb]{ .992,  .89,  .518}34.36 & \cellcolor[rgb]{ .992,  .894,  .518}34.69 & \cellcolor[rgb]{ .988,  .863,  .518}32.16 &       & \cellcolor[rgb]{ .98,  .824,  .518}99.07 & \cellcolor[rgb]{ .8,  .867,  .51}151.31 & \cellcolor[rgb]{ .976,  .8,  .518}90.08 & \cellcolor[rgb]{ .714,  .839,  .502}157.91 \\
    Tokyo & \cellcolor[rgb]{ .992,  .882,  .518}33.65 & \cellcolor[rgb]{ .98,  .831,  .518}29.60 & \cellcolor[rgb]{ .98,  .824,  .518}28.87 & \cellcolor[rgb]{ .984,  .855,  .518}31.56 &       & \cellcolor[rgb]{ .976,  .8,  .518}90.37 & \cellcolor[rgb]{ .906,  .894,  .514}143.32 & \cellcolor[rgb]{ .973,  .773,  .518}80.14 & \cellcolor[rgb]{ .737,  .847,  .506}156.34 \\
    Washington & \cellcolor[rgb]{ .976,  .918,  .518}37.97 & \cellcolor[rgb]{ .992,  .894,  .518}34.77 & \cellcolor[rgb]{ .992,  .886,  .518}34.16 & \cellcolor[rgb]{ .996,  .91,  .518}35.86 &       & \cellcolor[rgb]{ .984,  .847,  .518}108.45 & \cellcolor[rgb]{ .447,  .765,  .486}178.29 & \cellcolor[rgb]{ .976,  .808,  .518}93.76 & \cellcolor[rgb]{ .714,  .839,  .502}157.91 \bigstrut[b]\\
\hline  Average & \cellcolor[rgb]{ .929,  .902,  .514}40.49 & \cellcolor[rgb]{ .996,  .918,  .518}36.71 & \cellcolor[rgb]{ .996,  .914,  .518}36.39 & \cellcolor[rgb]{ .996,  .918,  .518}36.65 &       & \cellcolor[rgb]{ .984,  .847,  .518}108.03 & \cellcolor[rgb]{ .663,  .827,  .502}161.82 & \cellcolor[rgb]{ .98,  .82,  .518}97.33 & \cellcolor[rgb]{ .773,  .859,  .506}153.55 \bigstrut\\
    \hline
    \end{tabular}%
  \label{tab:G20_results}%
\end{table}%

\begin{table}[htbp]
  \centering
  \caption{Distributed message size for the Indo-ASEAN network of ground stations for constellations of 6 and 16 satellites. See caption of \Cref{tab:G20_results} for explanation of the notation. } 
  \begin{tabular}{|p{6em}|c|c|c|c|c|cc|cc|}
    \hline
    \textbf{IndoASEAN} & \multicolumn{4}{p{11em}}{6 satellites} & \multicolumn{1}{c}{} & \multicolumn{4}{p{11.29em}|}{16 satellites} \bigstrut\\
    \hline
    \multicolumn{1}{|r|}{} & \multicolumn{4}{p{11em}|}{ Message size [Mbit]} &       & \multicolumn{4}{p{11.29em}|}{ Message size [Mbit]} \bigstrut\\
\cline{2-5}\cline{7-10}    Location & \multicolumn{1}{p{2.75em}|}{SSO 1p/6s} & \multicolumn{1}{p{2.75em}|}{30deg 2p/3s} & \multicolumn{1}{p{2.75em}|}{30 deg 3p/2s} & \multicolumn{1}{p{2.75em}|}{30 deg 6p/1s} &       & \multicolumn{1}{p{2.75em}|}{SSO 1p/16s } & \multicolumn{1}{p{2.75em}|}{SSO 1p/16s ISL} & \multicolumn{1}{p{2.75em}|}{30 deg 1p/16s} & \multicolumn{1}{p{3.04em}|}{30deg 1p/16s ISL} \bigstrut\\
    \hline
    Singapore & \cellcolor[rgb]{ .957,  .69,  .518}10.13 & \cellcolor[rgb]{ .961,  .729,  .518}16.33 & \cellcolor[rgb]{ .961,  .725,  .518}15.84 & \cellcolor[rgb]{ .961,  .722,  .518}14.98 &       & \cellcolor[rgb]{ .961,  .729,  .518}28.36 & \cellcolor[rgb]{ .973,  .792,  .518}56.71 & \cellcolor[rgb]{ .969,  .757,  .518}40.46 & \cellcolor[rgb]{ .984,  .843,  .518}80.92 \bigstrut[t]\\
    Jakarta & \cellcolor[rgb]{ .988,  .867,  .518}37.13 & \cellcolor[rgb]{ .929,  .902,  .514}59.00 & \cellcolor[rgb]{ .925,  .902,  .514}60.34 & \cellcolor[rgb]{ .937,  .906,  .518}58.11 &       & \cellcolor[rgb]{ .992,  .882,  .518}98.43 & \cellcolor[rgb]{ .902,  .894,  .514}178.76 & \cellcolor[rgb]{ .933,  .902,  .514}159.49 & \cellcolor[rgb]{ .686,  .831,  .502}317.59 \\
    Delhi & \cellcolor[rgb]{ .98,  .918,  .518}49.49 & \cellcolor[rgb]{ .388,  .745,  .482}162.59 & \cellcolor[rgb]{ .412,  .753,  .486}158.37 & \cellcolor[rgb]{ .459,  .765,  .486}149.78 &       & \cellcolor[rgb]{ .918,  .898,  .514}170.08 & \cellcolor[rgb]{ .902,  .894,  .514}179.29 & \cellcolor[rgb]{ .553,  .792,  .494}404.61 & \cellcolor[rgb]{ .388,  .745,  .482}509.18 \\
    Mumbai & \cellcolor[rgb]{ .906,  .898,  .514}63.57 & \cellcolor[rgb]{ .533,  .788,  .494}135.02 & \cellcolor[rgb]{ .545,  .792,  .494}132.86 & \cellcolor[rgb]{ .549,  .792,  .494}132.16 &       & \cellcolor[rgb]{ .902,  .894,  .514}179.22 & \cellcolor[rgb]{ .902,  .894,  .514}179.29 & \cellcolor[rgb]{ .635,  .816,  .498}351.62 & \cellcolor[rgb]{ .392,  .749,  .486}509.18 \\
    Bengaluru & \cellcolor[rgb]{ .992,  .882,  .518}39.86 & \cellcolor[rgb]{ .886,  .89,  .514}67.50 & \cellcolor[rgb]{ .89,  .89,  .514}66.80 & \cellcolor[rgb]{ .89,  .89,  .514}66.76 &       & \cellcolor[rgb]{ .996,  .914,  .518}113.12 & \cellcolor[rgb]{ .902,  .894,  .514}179.29 & \cellcolor[rgb]{ .898,  .894,  .514}182.95 & \cellcolor[rgb]{ .616,  .812,  .498}364.19 \\
    Calcutta & \cellcolor[rgb]{ .992,  .886,  .518}40.37 & \cellcolor[rgb]{ .667,  .827,  .502}109.76 & \cellcolor[rgb]{ .686,  .831,  .502}105.97 & \cellcolor[rgb]{ .69,  .835,  .502}104.98 &       & \cellcolor[rgb]{ .996,  .922,  .518}119.41 & \cellcolor[rgb]{ .902,  .894,  .514}179.29 & \cellcolor[rgb]{ .737,  .847,  .506}286.56 & \cellcolor[rgb]{ .388,  .745,  .482}509.18 \\
    Kuala Lumpur & \cellcolor[rgb]{ .961,  .725,  .518}15.52 & \cellcolor[rgb]{ .969,  .757,  .518}20.51 & \cellcolor[rgb]{ .973,  .773,  .518}23.17 & \cellcolor[rgb]{ .973,  .78,  .518}23.93 &       & \cellcolor[rgb]{ .965,  .745,  .518}36.22 & \cellcolor[rgb]{ .98,  .824,  .518}71.95 & \cellcolor[rgb]{ .976,  .8,  .518}60.31 & \cellcolor[rgb]{ .992,  .922,  .518}120.62 \\
    Chennai & \cellcolor[rgb]{ .973,  .776,  .518}23.55 & \cellcolor[rgb]{ .996,  .914,  .518}44.36 & \cellcolor[rgb]{ .996,  .906,  .518}43.12 & \cellcolor[rgb]{ .996,  .902,  .518}42.47 &       & \cellcolor[rgb]{ .976,  .804,  .518}62.61 & \cellcolor[rgb]{ .988,  .918,  .518}124.36 & \cellcolor[rgb]{ .996,  .918,  .518}113.99 & \cellcolor[rgb]{ .831,  .875,  .51}226.32 \\
    Bangkok & \cellcolor[rgb]{ .98,  .816,  .518}29.75 & \cellcolor[rgb]{ .973,  .914,  .518}50.83 & \cellcolor[rgb]{ .969,  .914,  .518}52.05 & \cellcolor[rgb]{ .976,  .918,  .518}50.35 &       & \cellcolor[rgb]{ .984,  .839,  .518}78.42 & \cellcolor[rgb]{ .941,  .906,  .518}155.45 & \cellcolor[rgb]{ .965,  .914,  .518}138.20 & \cellcolor[rgb]{ .757,  .851,  .506}273.55 \\
    Ho Chi Minh & \cellcolor[rgb]{ .969,  .765,  .518}21.59 & \cellcolor[rgb]{ .984,  .843,  .518}33.96 & \cellcolor[rgb]{ .984,  .843,  .518}34.00 & \cellcolor[rgb]{ .984,  .839,  .518}33.33 &       & \cellcolor[rgb]{ .973,  .792,  .518}57.61 & \cellcolor[rgb]{ 1,  .922,  .518}115.22 & \cellcolor[rgb]{ .988,  .867,  .518}91.77 & \cellcolor[rgb]{ .894,  .894,  .514}183.54 \\
    Manila & \cellcolor[rgb]{ .973,  .788,  .518}25.36 & \cellcolor[rgb]{ .996,  .922,  .518}46.48 & \cellcolor[rgb]{ .992,  .894,  .518}41.81 & \cellcolor[rgb]{ .996,  .914,  .518}44.37 &       & \cellcolor[rgb]{ .98,  .82,  .518}69.80 & \cellcolor[rgb]{ .965,  .914,  .518}139.21 & \cellcolor[rgb]{ 1,  .922,  .518}116.90 & \cellcolor[rgb]{ .82,  .871,  .51}232.79 \\
    \hline
    Average & \cellcolor[rgb]{ .98,  .835,  .518}32.39 & \cellcolor[rgb]{ .886,  .89,  .514}67.85 & \cellcolor[rgb]{ .89,  .89,  .514}66.76 & \cellcolor[rgb]{ .898,  .894,  .514}65.57 &       & \cellcolor[rgb]{ .988,  .871,  .518}92.12 & \cellcolor[rgb]{ .961,  .91,  .518}141.71 & \cellcolor[rgb]{ .906,  .894,  .514}176.99 & \cellcolor[rgb]{ .71,  .839,  .502}302.46 \bigstrut[b]\\
    \hline
    \end{tabular}%
  \label{tab:IA_results}%
\end{table}%

\section{Discussion} 


The following sections discuss how these results impact the design of a constellation and the relevance of inter-satellite links.

\subsection{Constellation design}
Satellites in low inclination orbits pass up to 16 times per day over low latitude ground stations, while ground stations at high latitude typically only see satellites in high inclination orbits twice a day. Maximum latitudes in the G20 and Indo-ASEAN network are around 60 and 30 degrees, respectively. When restricted to a single orbital plane, maximising passes in the network is achieved by setting the inclination of the plane equal to the maximum latitude. This is illustrated in \Cref{fig:IA_single}. However, not all passes occur during nighttime. \Cref{fig:G20_single} shows that a noon-midnight sun-synchronous orbit, while having fewer passes overall, maximises useful accesses because of the long and consistent passes at nighttime it provides. Constellations with satellites spread over multiple orbital planes do not improve key sizes although the time between passes for individual ground stations may be reduced due to the improved temporal coverage \cite{Loarte2019}. Some ground stations consistently perform poorly due to clouds (modelled with a 0.1 longitude by 0.1 degree latitude resolution). While this emphasizes the need for choosing appropriate ground station locations, this is not necessarily problematic as any satellite-QKD network should interface with ground-based fibre networks that could cover the most cloudy areas.

\subsection{Inter-satellite links}
Results in \Cref{tab:G20_results} and \Cref{tab:IA_results} illustrate that inter-satellite links can provide a 20-100\% increase in message size for all ground stations for the two networks studied here under the embassy model assumption. It should be noted that for the chosen figure of merit there can be improvements in stored key even for the ground stations that originally have the most key. This is evident in the results for the Indo-ASEAN constellation, where the initial differences are very large, but not in the G20 constellation. Re-distribution is currently based on a simple algorithm that only copies keys from one satellite to the next etc. so further improvements in equalising key material can be envisaged. Whether ISL QKD is cost-effective was not considered within the scope of this work: a trade-off between simply launching more satellites versus increasing the complexity of individual satellites would be required. However, there is a number of satellites for a given network that makes optimum use of ISL, as the required shared key between satellites is determined by the size of the key that is to be re-distributed. Presently, the operational requirements of ISL do not lead to sub-optimal constellation choices, however more exotic constellation types may be derived that cannot feature ISLs.


\section{Conclusion} 

LEO constellations of trusted-node QKD satellites, continually performing QKD with the ground stations they fly over, can be used to provide low-latency, symmetric keys on demand between any two ground stations. They can do this by building a buffer of keys on board that can be quickly combined by an XOR operation and delivered to ground station via an RF (classical communications) relay satellite. The QKD satellites are best launched into SSO orbit or an orbit with an inclination equal to the latitude of the ground station farthest from the equator.

Inter-satellite links can improve the efficiency with which encryption keys are used, an average message size increase of 50-70\% in our examples, but whether this is cost-effective remains an open question. 

\section{Acknowledgments} 

Sergio Loarte performed the bulk of this work at the Centre for Quantum Technologies during an exchange stay. Hans Kuiper supervised Serio at TU Delft. With thanks to Abhijit Mitra and Sanat Biswas from IIIT Delhi who mooted the concept of an Indo ASEAN QKD network with us.\\ 

This work is partially supported by the National Research Foundation, Prime Minister’s Office, Singapore (under the Research Centres of Excellence programme and through Award No. NRF-CRP12-2013-02) and by the Singapore Ministry of Education (partly through the Academic Research Fund Tier 3 MOE2012-T3-1-009).\\

\section*{References}

\providecommand{\newblock}{}

\section*{Appendix A}
\label{appendixA}

\begin{table}[H]
\centering
\caption{\label{tab:miciusvalidation}Detailed comparison of Micius pass on 19th of December 2016}
\begin{indented}
\item[]\begin{tabular}{@{}llll}
\br
Parameter& Micius data & Model results & Units\\ \br
Closest approach & 645 & 635.7 & km\\
Sifted key rate at 1200 km & 1 & 1.2 & kbit/s\\
Sifted key rate at 645 km & 12 & 13.8 & kbit/s\\
Experiment duration  & 273  & 273 &s\\
Total detection events & 3,551,136 & 3,926,729 & bits \\
Sifted key size  & 1,671,072 & 1,963,364 & bits \\
Average Quantum Bit Error Rate & 1.1 & 1.2 & \%\\
Secret key size & 300,939 & 521,513 & bits\\
Average secret key rate  & 1102 & 1910 & bits/s\\
\end{tabular}
\end{indented}
\end{table}

\section*{Appendix B}
 \begin{figure}[H]
  \centering
  \includegraphics[width =\textwidth]{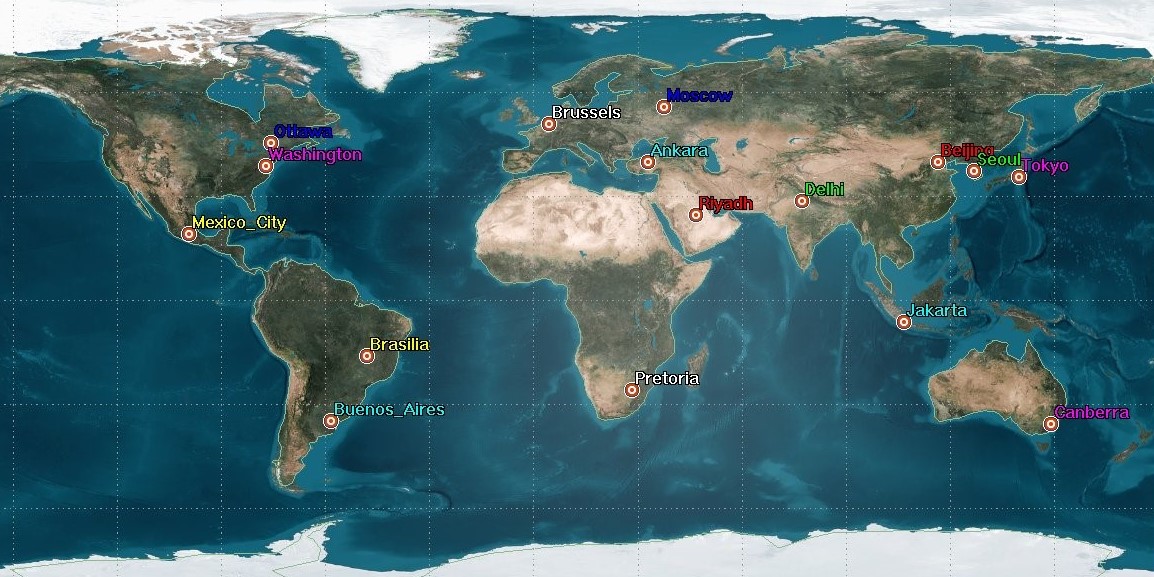}\\ \hspace{2cm}
  \caption{G20 ground stations}
  \label{fig:G20}
 \end{figure}
 
 \begin{figure}[H]
  \centering
  \includegraphics[width =\textwidth]{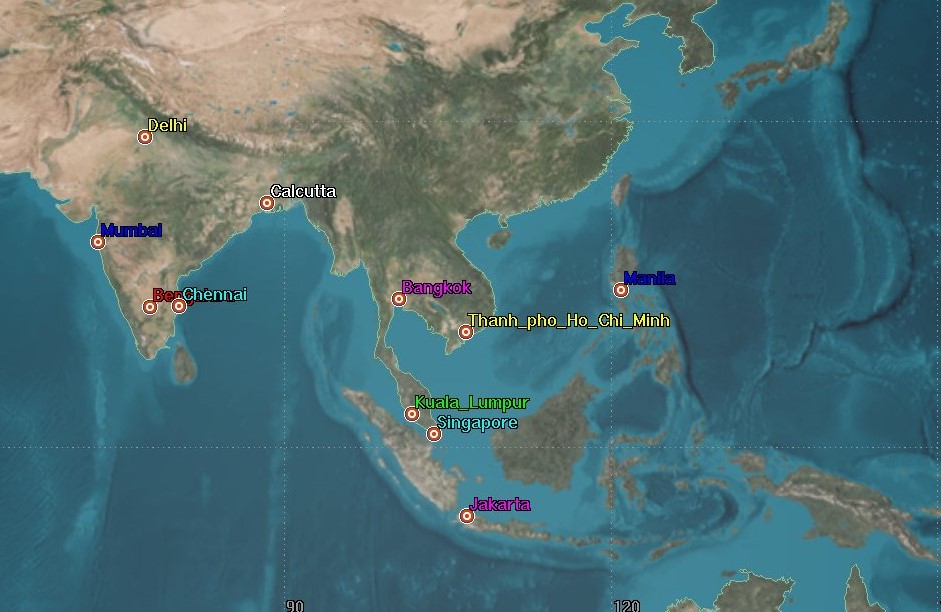}
  \caption{Indo-ASEAN ground stations}
  \label{fig:IndoASEAN}
 \end{figure}

\end{document}